\documentclass[letter]{aa}
\usepackage{graphicx}
\usepackage{amssymb}

\usepackage{txfonts}
\usepackage{color}
\usepackage{natbib}
\usepackage{tabularx}
\usepackage{multirow}

\begin{document}

\title{Finding short GRB remnants in globular clusters: the VHE gamma-ray source 
in Terzan~5}

\author{Wilfried F. Domainko\inst{1}}

\institute{Max-Planck-Institut f\"ur Kernphysik, P.O. Box 103980, D 69029
Heidelberg, Germany}

\offprints{\email{wilfried.domainko@mpi-hd.mpg.de}}

\date{}
 
\abstract
{Globular cluster are believed to boost the rate of compact binary mergers that may launch a certain type of cosmological gamma-ray burst (GRB). Therefore globular clusters appear to be potential sites to search for remnants of such GRBs.}
{The very-high-energy (VHE) gamma-ray source HESS~J1747-248 recently discovered in the direction of the Galactic globular cluster Terzan~5 is investigated for being a GRB remnant.}
{Signatures created by the ultrarelativistic outflow, the subrelativistic ejecta and the ionizing radiation of a short GRB are estimated for an expected age of such a remnant of t $\gtrsim 10^4$ years.}
{The kinetic energy of a short GRB could roughly be adequate for powering the VHE source in a hadronic scenario.
The age of the proposed remnant estimated from its extension possibly agrees with the occurrence of such events in the Galaxy. 
Subrelativistic merger ejecta could shock-heat the ambient medium.}
{Further VHE observations can probe for the presence of a break towards lower energies expected for particle acceleration in ultrarelativistic shocks. Deep X-ray observations would have the potential to examine whether there is any thermal plasma heated by the subrelativistic ejecta. The identification of a GRB remnant in our own Galaxy may also help to explore the effect of such a highly energetic event on the Earth.}
\keywords{ISM: supernova remnants -- Gamma-ray burst: general -- Galaxy:globular clusters: individual: Terzan~5}

\maketitle

\authorrunning{W. Domainko}
\titlerunning{A GRB remnant in Terzan~5}

\section{Introduction}

Gamma-ray bursts (GRBs) are enigmatic explosions, detected usually at cosmological
distances. They appear to come in two distinct flavors according to their duration \citep[see e.g.][for a review]{gehrels09}. Powerful explosions usually leave long-living remnants behind that potentially could be studied in the local Universe as relics of the original events.
 
Long bursts (duration longer than about 2 s) are generally considered to be launched by the death of a massive star and several structures in our Galaxy have been suggested to be the remnants of such events. These proposed remnants are in many cases linked to sources of very-high-energy (VHE, $>$100 GeV) gamma-ray emission. The VHE gamma-ray source HESS~J1303-631 \citep{aharonian05} has been suggested as the remnant of a long GRB that happened $\gtrsim 10^4$ years ago \citep{atoyan06}. Also the structure W49B has been suspected to be a remnant of a GRB \citep{ioka04}. Recently it has been argued that the population of unidentified TeV sources \citep[see e.g.][]{aharonian08} may be dominated by GRB/Hypernova remnants \citep{ioka09}.

For short GRBs (duration shorter than 2 s), compact binary mergers as central engine have been identified as the preferred scenario \citep[e.g.][]{gehrels05,lee05}. Such compact binaries may be efficiently created in globular clusters since these environments feature high densities of very old stars in their cores. Indeed in the globular cluster M~15 already a close system that consists of two neutron stars (NS) has been discovered \citep{anderson90}. Based on these facts, it has been argued that a considerable fraction of all NSNS binaries are formed in globular clusters \citep{grindlay06}. It has been claimed that the rate of short bursts in the local Universe is dominated by mergers of dynamically formed compact binaries in globular clusters \citep{salvaterra08,guetta09}. Therefore, globular clusters might be the prime environment for seeking remnants of short GRBs. Potential signatures of remnants of compact binary mergers have already been discussed \citep[see][]{domainko05,domainko08}.

The H.E.S.S. collaboration has very recently reported the detection of a VHE gamma-ray source HESS~J1747-248 in the direction of the galactic globular cluster Terzan~5 \citep{hesspaper} with a flux above 440 GeV of (1.2$\pm$0.3)$\times$10$^{-12}$~cm$^{-2}\,$s$^{-1}$. The source is extended with intrinsic extension of 9$^\prime$.6$ \pm $2$^\prime$.4, and there appears to be an offset from the cluster core position by about 4$^\prime$.0$ \pm $1$^\prime$.9. Terzan~5 is located at a distance of 5.9 kpc \citep{ferraro2009}, at RA(J2000)~17$^\mathrm{h}$48$^\mathrm{m}$04$^\mathrm{s}$.85 and Dec~$-24^{\circ}$46$^\prime$44$^{\prime\prime}$.6 
(l 3.8$^{\circ}$, b 1.7$^{\circ}$), and exhibits a core radius of 0.15' a half-mass radius of 0.52' and a tidal radius of 4.6' \citep{lanzoni2010}. Terzan~5 is the globular cluster with the highest expected rates of close stellar encounters \citep{pooley06} and
with the largest population of millisecond pulsars (33) discovered up to now \citep{ransom08}. Gamma rays of likely magnetospheric origin produced by millisecond pulsars have been detected in the GeV range in Terzan~5 \citep{kong10,abdo10}. Diffuse X-ray emission extending beyond the half-mass radius has also been reported \citep{eger2010}. In the vicinity of Terzan~5 several structures in the radio band have been found \citep{clapson2011}. Terzan~ 5 is a predicted VHE gamma-ray emitter where energetic electrons produced by the large population of millisecond pulsars up-scatter stellar photons to gamma-ray energies \citep{venter2009,bednarek2007}.
In contrast to these models here VHE gamma-ray emission originating from collisions of hadronic cosmic rays with ambient target nuclei and subsequent $\pi^0$ decay is explored. A short GRB is adopted as the accelerator of the cosmic rays.

The paper is organized as follows: In Sect. \ref{sec:hadronic} potential indications for a hadronic VHE gamma-ray production in Terzan~5 are discussed, in Sect. \ref{sec:ultra_rel} signatures left behind by the ultrarelativistic outflow of an ancient GRB in Terzan~5 are assessed, in Sect. \ref{sec:sub_rel} signatures related to the subrelativistic ejecta are explored, and in Sect. \ref{sec:ionization} potential traces of ionizing radiation emitted by the GRB are investigated. 

\section{Potential indications for a hadronic scenario}\label{sec:hadronic}

Some of the properties of the VHE gamma-ray source that were detected in the direction of Terzan~5 appear to challenge leptonic models for the gamma-ray production. In particular, 
the extension, indication for an offset of the source from the globular cluster core and a power-law spectrum are not self-evident for such a scenario.
The intensity of leptonic inverse-Compton (IC) radiation scales linearly with the energy density 
of the target photon field. For Terzan~5 the energy density of the stellar photon field 
drops from about 1000 eV/cm$^{-3}$ in the core region to 40 eV/cm$^{-3}$ 
at the half-mass radius \citep{venter2009} to about a few eV/cm$^{-3}$ at the extension 
of the VHE source. Consequently a very centrally peaked source centered on 
the GC would be expected, which does not seem to be supported by the H.E.S.S. 
observations. IC emission in the VHE range should be accompanied by synchrotron emission in the X-ray band. Diffuse X-ray emission centered on Terzan~5 of possible synchrotron origin has indeed been discovered \citep{eger2010} but the potential offset of the VHE gamma-ray emission from the peaks of the X-ray emission and radiation field challenges a leptonic scenario.
Furthermore, since the optical to near-infrared starlight 
photon field should be up-scattered by the very-high energy electrons, 
Klein-Nishina (KN) suppression of the IC process should be significant 
at multiple TeV energies causing a steepening in the VHE gamma-ray spectrum. 
For a target stellar photon field with mean temperature of 4500 K \citep{venter2009} 
and electron energies of 10 TeV the KN suppression factor is already 
about 0.025 \citep{coppi1990}, so the VHE gamma-ray spectrum should steepen 
well before this energy. In comparison, the observed spectrum may follow a straight power law, but this result is influenced by limited statistics. To account for the aforementioned arguments, as an alternative to an IC scenario, hadronic gamma-ray production is explored in this paper as the origin of the VHE source.

\section{Ultrarelativistic outflow}\label{sec:ultra_rel}

\subsection{Energetics}

GRBs are generally believed to be caused by a pair of ultrarelativistic jets that are ejected from the central engine. Relativistic shock waves accelerate all particles from the incoming plasma to relativistic energies \citep{blandford76}, thus a substantial fraction of the initial energy of the relativistic blast wave is transferred into cosmic rays \citep[see][for the case of a GRB remnant]{atoyan06}. 
Therefore, for such a scenario, the energy in cosmic rays is a measure for the kinetic energy of the relativistic outflow.

From the luminosity of the VHE gamma-ray source the energy in hadronic cosmic rays can be estimated if the density of target material is assumed. At the location of Terzan~5 the density of target material should be on the order of $n \approx 0.1$~cm$^{-3}$ \citep{clapson2011}. Additionally, to constrain the total energy in hadronic cosmic rays for the entire relevant cosmic ray energy range above about 1 GeV 
a spectral index for the region below the range that can be probed with H.E.S.S. has to be assumed.
If a cosmic ray spectral index of 2.0 is adopted below 5 TeV, the cosmic ray energy that produces gamma rays at the energy threshold of 440 GeV of the H.E.S.S. measurements, the total energy in hadronic cosmic rays would be $E_\mathrm{CR} \approx 10^{51} (n/0.1 \mathrm{cm}^{-3})^{-1}$ ergs \citep{hesspaper}.

Particles accelerated by a relativistic shock wave will not feature a single power-law spectrum but should have a break towards low energies where the break energy $E_\mathrm{br}$ is given by the bulk Lorenz factor $\Gamma$ of the relativistic shock \citep{blandford76,katz94} according to $E_\mathrm{br} \sim m_\mathrm{p} c^2 \Gamma^2 /2$ with $m_\mathrm{p}$ the mass of the particle and $c$ the speed of light. If a proton energy of 5 TeV, which produces gamma rays at the H.E.S.S. threshold, is adopted as the break energy, then the Lorenz factor at the time when most particles are accelerated would be $\lesssim$ 100. Lower break energies would result in lower Lorenz factors. A spectral break towards lower particle energies would reduce the $E_\mathrm{CR}$ needed to explain the VHE source depending on the break energy by up to a factor of 2 \citep{atoyan06}. Since ultrarelativistic blast waves are expected to transfer a substantial part of their kinetic energy into cosmic rays the energetics of short GRBs could be roughly adequate for the observed VHE source if a ratio of prompt electromagnetic energy release to kinetic energy of 0.1 - 0.01 \citep{nakar07} is assumed. Extending VHE observations to lower energies with the planned CTA array \citep{cta2010} could probe for a break in the gamma-ray spectrum of this source to test this scenario.

\subsection{Age of the remnant and rate of short GRBs}

After the acceleration, cosmic rays will diffuse away from the location of the GRB and will therefore form extended, center-filled gamma-ray sources \citep{atoyan06}.
The age of a GRB remnant would in such a case be given by the diffusive propagation time of cosmic rays to the extension of the source. 
For the VHE source in Terzan~5 an age of the remnant of $10^3 (D/ 10^{28}$~cm$^2$s$^{-1})^{-1}$~years would be found \citep{hesspaper}, where $D$ is the uncertain diffusion coefficient here compared to the value estimated for 5 TeV protons in the galactic disk of $10^{28}$ cm$^2$s$^{-1}$ \citep{atoyan06}.

The age obtained for the potential GRB remnant at Terzan~5 can be compared to the rate of compact binary mergers in the Galaxy and to the rate of short GRBs in the local Universe. From field NSNS binaries a galactic merger rate to one event per (0.5 - 7)$\times 10^4$ years is found \citep{kalogera04}. For merger induced bursts which are formed in globular clusters a local rate of 20 - 90 events per Gpc$^{-3}$yr$^{-1}$ \citep{salvaterra08} or $\sim$4~Gpc$^{-3}$~yr$^{-1}$ \citep{guetta09} has been estimated. With a density of Milky way-type galaxies in the local Universe of 0.01 galaxies per Mpc$^{-3}$ \citep{cole01} this results in a rate of short bursts per galaxy of about one event per (0.1 - 0.5)$\times 10^4$ $(f_\mathrm{b}^{-1}/100)^{-1}$ years or 2.5$\times 10^4$ $(f_\mathrm{b}^{-1}/100)^{-1}$ years. Here $f_\mathrm{b}$ is the beaming factor of short bursts, uncertain in the range of 1 $\ll$ $f_\mathrm{b}^{-1}$ $<$ 100 \citep{nakar07}. It appears that the age of the potential GRB remnant would be roughly comparable to the rate of short bursts in the Milky Way if the diffusion coefficient $D$ were below $10^{27}$~cm$^{2}\,$s$^{-1}$ in the TeV range. Slowed down diffusion at the location of Terzan~5 might be a reasonable assumption since X-ray observations demonstrated that the plasma there is highly turbulent \citep{yao2007,crocker2010}. To conclude, for possible diffusion coefficients the age estimated for the hypothetical remnant could be roughly comparable to the merger occurrence.

\subsection{Potential multiwavelength signatures}

Nonthermal diffuse X-ray emission from Terzan~5 extending beyond the half-mass radius has also been detected \citep{eger2010}. The surface brightness of this X-ray emission roughly follows the surface brightness of the stellar component of the globular cluster, which could indicate an IC origin.
In the framework of a GRB remnant scenario 
these diffuse X-rays might be emitted by primary electrons accelerated by the ultrarelativistic blast wave. Primary electrons would have cooled down below GeV energies since their production and therefore would not emit synchrotron radiation in the GHz band \citep{atoyan06} but might still be energetic enough to scatter the intense stellar radiation field in the globular cluster to the X-ray range. For such a scenario an energy in electrons of $5 \times 10^{49}$~($u_\mathrm{rad}$/40~eV~cm$^{-3}$)~ergs would be required \citep{eger2010} with $u_\mathrm{rad}$ being the mean energy density in the stellar radiation field in the diffuse X-ray source. If this situation is indeed realized in Terzan~5, then a comparison to the energy in hadronic cosmic rays required to explain the VHE gamma-ray emission can be made. Here a quite large electron-to-proton ratio of $\approx 0.1$ even after the electrons have cooled down below GeV energies would be found.
Observations in the hard X-ray regime to extend the spectrum of the X-ray source might shed more light on its origin.

\section{Subrelativistic ejecta}\label{sec:sub_rel}

\subsection{Pressure driven remnant}

During the merger of compact binaries, a small fraction of the NS matter ($10^{-4}$ - 0.1 M$_\odot$) is dynamically ejected from the system with subrelativistic velocities \citep[e.g.][]{ruffert96,lee99,rosswog05}. When assuming a velocity of the ejected material of $c/3$ ($c$ speed of light), its kinetic energy would be 10$^{49}$ - 10$^{52}$ ergs. After the ejecta has displaced a comparable mass of ambient medium ($n \approx 0.1$~cm$^{-3}$), most of the kinetic energy is converted into thermal energy \citep[e.g.][]{dorfi1990} and the remnant expands farther in the pressure-driven phase \citep{domainko05,domainko08}. It should, however, be noted that a large population of hadronic cosmic rays produced in the ultrarelativistic outflow may alter the evolution of the non-relativistic blast wave \citep[e.g.][]{chugai2011}. In the pressure-driven stage, the remnant is filled with hot thermal plasma that might be observable in thermal X-ray emission. Cooling of the thermal plasma appears to be unimportant in a remnant of the proposed age of a few times 10$^4$ years, since material with a density of 0.1 cm$^{-1}$ and temperature of 10$^7$ (10$^6$) K would cool on a timescale of about $10^{7}$ ($3 \times 10^{5})$ years when X-ray line emission is considered \citep{sarazin1986}. Not detecting an additional thermal component above the galactic diffuse X-ray emission and the extended nonthermal X-ray source with \textit{Chandra} 
\citep{eger2010} would favor a content of hot thermal plasma in the potential remnant at the lower end of model predictions. A pressure driven remnant with energetics of 10$^{49}$ ergs and an assumed age of 10$^4$ years would extend to a radius of about 10 pc corresponding to an angular radius of 6$^\prime$ at a distance of 5.9~kpc. It would be filled with thermal plasma of about $ T \approx 10^{7}$~K for a mean density of $n \approx 0.1$~cm$^{-3}$, which should radiate $4 \times 10^{33}\,(T/10^{7}\mathrm{K})^{-0.6}\, (n/0.1 \mathrm{cm}^{-3})^2$ ergs s$^{-1}$ in X-rays, corresponding to 10$^{-12}\,(T/10^{7}\mathrm{K})^{-0.6}\, (n/0.1 \mathrm{cm}^{-3})^2$ ergs cm$^{-2}$ s$^{-1}$ at Earth \citep[X-ray emissivity from][]{sarazin1986}. Such an extended and faint structure might have been missed by the present X-ray observations due to the limited field of view of \textit{Chandra} and the limited sensitivity of \textit{ROSAT}. Deep future X-ray observations with the existing \textit{XMM} and the planned \textit{IXO} mission \citep{barcons2011} would have the potential to probe the presence of diffuse thermal X-ray emission in the direction of Terzan~5.

\subsection{Signatures from the ejecta}\label{sec:ejecta}

The ejecta of compact binary mergers may consist of r-process nuclei \citep[e.g.][]{lattimer74,ruffert97,freiburghaus99}. Many of these heavy nuclei are radioactive and emit hard X-ray and soft gamma-ray line emission during their decay \citep[e.g.][]{qian99,domainko05,domainko08}. The strength of these nuclear lines depend on the decay properties of the respective nuclei \citep{qian99}. Promising nuclei in terms of detectability of the gamma-ray line emission comprise a half-life time $\tau$ comparable to the age of the remnant. The estimated flux values in the table are given for a mass of the respective nuclei of 10$^{-5}$ M$_\odot$, an age of the remnant of 10$^4$, years and a distance to Terzan~5 of 5.9 kpc. Even for such an advantageous case, the line brightness would not exceed 10$^{-8}$~$\gamma$~cm$^{-2}$~s$^{-1}$, and detection would be challenging for the next generation of hard X-ray / soft gamma-ray instruments \citep[e.g.][]{knoedlseder2009}.

\begin{table}[h]\label{tab:lines}
\centering

\vspace{0.2cm}
\begin{tabular}{lcccccc}

\hline
r-process &   &   $\tau$   &  &  E$_{\gamma}$   &  &  F$_{\gamma}$   \\
nucleus   &  & [10$^3$ yr] &  & [keV] &  &  [10$^{-7}$ $\gamma$ cm$^{-2}$ s$^{-1}$]   \\
\hline

$^{226}$Ra & & 2.31 & & 609 & & 0.01 \\
$^{229}$Th & & 10.6 & & 40.0 & & 0.04 \\
$^{243}$Am & & 7.37 & & 74.7  & & 0.09 \\
\hline

\end{tabular}
\caption{Properties of gamma ray lines in a potential merger remnant in Terzan~5.}
\end{table}

\section{Ionizing radiation}\label{sec:ionization}

\subsection{Astrophysical signatures}

GRBs and their afterglows exhibit strong ionizing radiation which should alter the equilibrium state of the surrounding interstellar medium (ISM). Signatures of such an ionizing event should be observable for several 10$^4$ years in form of high-ionization lines \citep{perner2000}. Terzan~5 is, however, located in a direction close to the galactic center (l~3.8$^{\circ}$, b~1.7$^{\circ}$), hence affected by severe galactic absorption \citep[$A_\mathrm{V} = 7.72$,][]{ortolani1996}, which may challenge the detection of such emission lines.

\subsection{Terrestrial signatures}

The ionizing radiation of a GRB that is beamed towards the Earth may also alter the state of the Earth's atmosphere \citep{fishman88}, and Galactic GRBs should produce a substantial amount of NO$_{\mathrm{y}}$ there \citep{melott04}. This can be deposited as nitrates in the polar ice caps \citep[e.g.][]{thomas05,ejzak07}, which represent an archive of the atmospheric conditions for $7.4 \times 10^5$ years \citep{augustin04}. Similar signatures of nearby, Galactic supernovae have already been claimed \citep[e.g.][]{motizuki09}. Here the possibility of such a signature originating from a GRB in Terzan 5 is appraised. Considerations in this section are less well founded than discussions in previous sections. An evaluation of different terrestrial NO$_{\mathrm{y}}$ signatures can be found in \citet{harfoot2007}.
 
To assess NO$_{\mathrm{y}}$ deposition from a GRB in Terzan~5, the solar proton event that occurred during August 1972 and which is imprinted in polar ice \citep{zeller86,shea06} is used as a reference. This specific event produced $3.6\times 10^{33}$ molecules NO$_{\mathrm{y}}$ in the middle atmosphere \citep{jackman05}. 
It is found that a fluence of about $10^5$ ergs/cm$^2$ of ionizing radiation is needed for a comparable NO$_{\mathrm{y}}$ enhancement if it is adopted that 35 eV are necessary per ionization \citep{gehrels03} and that each ionization releases 1.25 N atoms, which rapidly produce NO$_{\mathrm{y}}$ \citep{jackman05}.
A GRB at an adopted distance of Terzan~5 of 5.9 kpc would have to release an isotropic fluence comparable to typical short bursts of several 10$^{50}$~ergs in ionizing radiation to produce enough nitrates to leave significant signatures in polar ice cores. Thus such an event could indeed have left traces in polar ice but only on the level of more frequent solar proton events. However, since GRBs are beamed with a beaming factor $f_\mathrm{b}$ in the range of 1 $\ll$ $f_\mathrm{b}^{-1}$ $<$ 100, the probability that the prompt GRB emission has hit the Earth would only be a few percent.

Along with generating nitrates by X-rays and soft gamma-rays, photons from the GRB with energies $>$ 10 MeV would produce long-living radionuclides like $^{10}$Be (half-life time 1.5 Myr) in the Earth atmosphere \citep{thorsett95}, which will after rain-out also be deposited in polar ice caps. Such high-energy emission has indeed been observed from some short GRBs \citep[e.g.][]{abdo10b}. In this framework it is interesting to note that two possibly cosmogenic $^{10}$Be anomalies have been found in antarctic ice cores \citep{raisbeck87}. However, the detectability of such signatures from a potential short GRB in Terzan~5 will again be limited by the beaming of the high-energy radiation. 

\section{Summary}

The VHE gamma-ray source discovered in the vicinity of Terzan~5 appears to be compatible with being the remnant of a short GRB induced by the merger of two compact stars. An incontrovertible proof is missing, but also no contradictory facts have been 
found for this scenario. Further observations in the X-ray and gamma-ray regime may help to strengthen the evidence for a GRB remnant in Terzan~5. In particular X-ray observations could probe the presence of thermal plasma heated by subrelativistic merger ejecta. The identification of a GRB remnant in our own Galaxy would provide important constraints on the GRB rate in the very local universe and may also help to explore the effect of such a highly energetic event on the Earth.


\begin{acknowledgements}

The author acknowledge support from his host institution. The author want to thank A.-C. Clapson and M. Ruffert for many enlightening discussions.

\end{acknowledgements}


\end{document}